\documentclass[graybox]{svmult}

\usepackage{mathptmx}       
\usepackage{helvet}         
\usepackage{courier}        
\usepackage{type1cm}        

\usepackage{makeidx}         
\usepackage{graphicx}        
\usepackage{multicol}        
\usepackage[bottom]{footmisc}

\makeindex

\begin{document}

\title*{Singular Problems for Integro-Differential  Equations in Dynamic
Insurance  Models}
\author{Tatiana Belkina, Nadezhda Konyukhova, and Sergey Kurochkin}
\authorrunning{T.~Belkina et al.}
\institute{T.~Belkina \at Central Economics and Mathematics
Institute of RAS, Nakhimovskii pr. 47, Moscow, Russia \\
\email{tbel@cemi.rssi.ru} \and N.~Konyukhova,  S.~Kurochkin \at
Dorodnicyn Computing Centre of RAS, ul. Vavilova 40, Moscow,
Russia \\ \email{nadja@ccas.ru; kuroch@ccas.ru}
\and S.~Pinelas et al. (eds.), \textit{Differential and Difference Equations with Applications}, Springer \\
Proceedings in Mathematics \& Statistics 47, DOI 10.1007/978-1-4614-7333-6-3 \\
}

\maketitle

\abstract*{ A second order linear integro-differential equation
with  Volterra integral operator and strong singularities at the
endpoints (zero and infinity) is considered.  Under limit
conditions at the singular points, and some natural assumptions,
the problem is a singular initial problem with limit normalizing
conditions at infinity. An existence and uniqueness theorem is
proved and asymptotic representations of the solution are given.
A numerical algorithm for evaluating the solution is proposed,
calculations and their interpretation are discussed. The main
singular problem under study describes the survival (non-ruin)
probability of an insurance company on infinite time interval
(as a function of initial surplus) in the Cram\'er-Lundberg
dynamic insurance model with an exponential claim size
distribution and certain company's strategy at the financial
market assuming investment of a fixed part of the surplus
(capital) into risky assets (shares) and the rest of it into a
risk free asset (bank deposit). Accompanying "degenerate"
problems are also considered that have an independent meaning in
risk theory.}

\abstract{ A second order linear integro-differential equation
with  Volterra integral operator and strong singularities at the
endpoints (zero and infinity) is considered.  Under limit
conditions at the singular points, and some natural assumptions,
the problem is a singular initial problem with limit normalizing
conditions at infinity. An existence and uniqueness theorem is
proved and asymptotic representations of the solution are given.
A numerical algorithm for evaluating the solution is proposed,
calculations and their interpretation are discussed. The main
singular problem under study describes the survival (non-ruin)
probability of an insurance company on infinite time interval
(as a function of initial surplus) in the Cram\'er-Lundberg
dynamic insurance model with an exponential claim size
distribution and certain company's strategy at the financial
market assuming investment of a fixed part of the surplus
(capital) into risky assets (shares) and the rest of it into a
risk free asset (bank deposit). Accompanying "degenerate"
problems are also considered that have an independent meaning in
risk theory.}
\section{Introduction}
The important problem concerning the application of financial
instruments in order to reduce insurance risks has been
extensively studied  in recent years (see, e.g.,
\cite{kony_AzMul}, \cite{kony_BelKonKur1},
\cite{kony_BelKonKur2}, and references therein). In particular in
\cite{kony_BelKonKur1}, \cite{kony_BelKonKur2} the optimal
investing strategy is studied  for risky and risk-free assets
in Cram\'er-Lundberg (C.-L.) model with budget constraint, i.e.,
without borrowing.

This paper complements and revises some results of
\cite{kony_BelKonKur2}. The  parametric singular initial problem
(SIP) for an integro-differential equation (IDE) considered here
is a part of the optimization problem stated and analyzed in
\cite{kony_BelKonKur1}, \cite{kony_BelKonKur2}: the solution of
this SIP gives the survival probability corresponding to the
optimal strategy when the initial surplus values are small
enough. The singular problem  under study is also interesting
both as an independent mathematical problem and for the models in
risk theory. We give more complete and rigorous analysis of this
problem in comparison with \cite{kony_BelKonKur2} and add some
new "degenerate" problems having independent meaning in risk
theory. Some new numerical results are also discussed.

The paper is organized as follows. In Sect. 2 we set the main
mathematical problem and formulate the main results concerning
solvability of this problem and the solution  behavior; we
describe also two "degenerate" problems (when some parameters in
the IDE are equal to zero) and discuss their exact solutions. In
Sect. 3 we give a rather brief description of the mathematical
model for which the problem in question arises (for detailed
history, models' description and derivation of the IDE studied
here, see \cite{kony_BelKonKur1}, \cite{kony_BelKonKur2}). In
Sect. 4 we describe our approach to the problem and give brief
proofs of main results (for some assertions, we omit the proofs
since they are given in \cite{kony_BelKonKur2}). In Sect. 5 we
study an accompanying singular problem for capital stock model
(the third "degenerate" problem); the results of this section
are completely new. Numerical results and their interpretation
are given in Sect. 6.
\section{ Singular Problems for IDEs and Their Solvability}
\label{sec:1}
\subsection{ Main Problem}
\label{subsec:1.1}
The main singular problem under consideration
has the form:
$$(b^2/2) u^2 \varphi^{\prime\prime}(u) + (au+c)
\varphi^{\prime}(u) - \lambda\varphi(u) + $$
\begin{equation}\label{kony_IDE_pr}
 + (\lambda/m) \int_0^u {\varphi(u-x)\exp{(-x/m)}
 dx}=0, \qquad 0<u<\infty,
\end{equation}
\begin{equation}\label{kony_IDE_cond_0_1_pr}
\{|\lim_{u\to+0}{\varphi(u)}|, |\lim_{u\to+0}
{\varphi^{\prime}(u)}|\}<\infty, \quad
\lim_{u\to+0}{[c\varphi^\prime(u) - \lambda
\varphi(u)]=0},
\end{equation}
\begin{equation}\label{kony_IDE_restr_pr}
0\le\varphi(u)\le 1, \qquad  u\in {\textbf{R}}_+,
\end{equation}
\begin{equation}\label{kony_IDE_cond_inf_pr}
\lim_{u\to\infty}\varphi(u)=1, \qquad \lim_{u\to\infty}
\varphi^\prime(u)=0.
\end{equation}
Here in general all the parameters $a$, $b$, $c$, $\lambda$, $m$
are real positive numbers.

The second limit condition at zero is a corollary of the first
one and IDE (\ref{kony_IDE_pr}) itself. For this IDE,
conditions (\ref{kony_IDE_cond_0_1_pr})
imply $\lim_{u\to+0}{[u^2 \varphi^{\prime\prime}(u)}]=0$ providing a
degeneracy of the IDE (\ref{kony_IDE_pr}) as $u\to+0$: any
solution $\varphi(u)$ to the singular problem without initial
data (\ref{kony_IDE_pr}), (\ref{kony_IDE_cond_0_1_pr}) must
satisfy IDE (\ref{kony_IDE_pr}) up to the singular point $u=0$.

The "truncated" problem
(\ref{kony_IDE_pr})-(\ref{kony_IDE_restr_pr}) (constrained
singular problem) always has the trivial solution
$\varphi(u)\equiv 0$. A nontrivial solution is singled out by
the additional limit conditions at infinity
(\ref{kony_IDE_cond_inf_pr}).

In what follows we use notation
\begin{equation}\label{int_exp_F_pr}
(J_m\varphi)(u)=\frac{1}{m}\int_0^u {\varphi(u-x)\exp{(-x/m)}
dx}=\frac{1}{m}\int_0^u{\varphi(s)\exp{(-(u-s)/m)}ds}
\end{equation}
where  $J_m$ is a Volterra integral operator, $J_m: C[0, \infty)\to C[0, \infty)$,
$C[0, \infty)$ is the linear space of continuous functions
defined and bounded on ${\textbf{R}}_+$.

For IDE (\ref{kony_IDE_pr}), the entire singular problem on
${\textbf{R}}_+$ was neither posed nor studied before
\cite{kony_BelKonKur2} and the present paper.
\subsection{Formulation of the Main Results}
\label{subsec:1.2}
The problem (\ref{kony_IDE_pr})-(\ref{kony_IDE_cond_inf_pr}) may be
rewritten in the equivalent parametrized form:
\begin{equation}\label{kony_1intdiff_exp_pr}
(b^2/2)u^2\varphi^{\prime\prime}(u)+(au+c)\varphi^{\prime}(u)
-\lambda[\varphi(u)-(J_m\varphi)(u)]=0, \quad u\in
{\textbf{R}}_+,
\end{equation}
\begin{equation}\label{kony_TrebDerivs_pr}
\lim_{u\to+0}{\varphi(u)}=C_0,  \qquad
\lim_{u\to+0}{\varphi^{\prime}(u)}=\lambda C_0/c,
\end{equation}
\begin{equation}\label{kony_IDE_restr_pr1}
0\le\varphi(u)\le 1, \qquad  u\in {\textbf{R}}_+,
\end{equation}
\begin{equation}\label{kony_IDE_cond_inf_pr1}
\lim_{u\to\infty}\varphi(u)=1, \qquad \lim_{u\to\infty}
\varphi^\prime(u)=0.
\end{equation}
Here $C_0$ is an unknown parameter whose value must be defined.
\begin{lemma}
\label{l:1}
For  IDE {\rm(\ref{kony_1intdiff_exp_pr})}, let the values $a$, $b$, $c$,
$\lambda$, $m$ be fixed with $b\not =0$, $c>0$, $\lambda\not=0$, $m>0$,
$a\in {\textbf{\textrm{R}}}$. Then for any fixed $C_0\in{{\textbf{\textrm{R}}}}$
the IDE SIP {\rm(\ref{kony_1intdiff_exp_pr})}, {\rm(\ref{kony_TrebDerivs_pr})}
is equivalent to the following singular Cauchy problem (SCP) for ODE:
\begin{equation}\label{kony_diffur_pr}
\begin{array}{c}
(b^2/2) u^2 \varphi^{\prime\prime\prime}(u) + \left [c +
(b^2 + a) u +  b^2 u^2/(2m)\right]\varphi^{\prime\prime}(u) + \\
\\ + \left (a - \lambda + c/m + a u/m \right) \varphi^\prime(u)=0,
\quad 0<u<\infty,
\end{array}
\end{equation}
\begin{equation}\label{kony_cond_0_C0_d_pr}
\begin{array}{c}
\lim_{u\to+0}{\varphi(u)} = C_0,  \quad
\lim_{u\to+0}{\varphi^{\prime}(u)}=\lambda C_0/c,
\\
\\
\lim_{u\to+0}{\varphi^{\prime\prime}(u)}=(\lambda-a - c/m)
\lambda C_0/c^2.
\end{array}
\end{equation}
There exists a unique solution $\varphi(u,C_0)$ to SCP
{\rm{(\ref{kony_diffur_pr}), (\ref{kony_cond_0_C0_d_pr})}}
(therefore also to the equivalent IDE SIP
{\rm(\ref{kony_1intdiff_exp_pr})},
{\rm(\ref{kony_TrebDerivs_pr})}); for small $u$, this solution
is represented by the asymptotic power series
\begin{equation}\label{kony_asymp00_pr}
\varphi(u,C_0) \sim C_0\left[1+\frac{\lambda}{c}\left(u+
\sum_{k=2}^{\infty}D_k u^k/k \right)\right], \qquad u\sim +0,
\end{equation}
where coefficients $D_k$ are independent of $C_0$ and may be
found by formal substitution of series
{\rm(\ref{kony_asymp00_pr})} into ODE
{\rm(\ref{kony_diffur_pr})}, namely from the recurrence
relations
\begin{equation}\label{kony_1D_2_pr}
D_2=-[(a-\lambda)/c+ 1/m],
\end{equation}
\begin{equation}\label{kony_1D_3_pr}
D_3=- [D_2(b^2+2a-\lambda +c/m)+a/m]/(2c),
\end{equation}
\begin{equation}\label{kony_1D_k_pr}
\begin{array}{c}
D_k= - \{D_{k-1}[(k-1)(k-2)b^2/2+(k-1)a-\lambda+c/m]+
\\
\\
+ D_{k-2}[(k-3)b^2/2+a]/m\}/[c(k-1)], \quad k=4,5,\ldots.
\end{array}
\end{equation}
\end{lemma}
\begin{theorem}
\label{t:1}
For  IDE {\rm{(\ref{kony_IDE_pr})}}, let
all the parameters $a$, $b$, $c$, $\lambda$, $m$ be fixed
positive numbers and let the inequality
\begin{equation}\label{kony_r_0_pr1}
2a/b^2>1
\end{equation}
be fulfilled. Then the following statements are valid:
\begin{enumerate}
\item[1.]
There exists a unique solution $\varphi(u)$ of the input
singular linear IDE problem
{\rm{(\ref{kony_IDE_pr})-(\ref{kony_IDE_cond_inf_pr})}} and it is a
smooth (infinitely differentiable) monotone nondecreasing on
${\textbf{R}_+}$ function.
\item[2.] The function $\varphi(u)$ can be obtained
as the solution $\varphi(u, C_0)$ of IDE SIP
{\rm(\ref{kony_1intdiff_exp_pr})},
{\rm(\ref{kony_TrebDerivs_pr})}, namely by solving the
equivalent ODE SCP {\rm{(\ref{kony_diffur_pr}),
(\ref{kony_cond_0_C0_d_pr})}} where the value $C_0=\widetilde
C_0$ must be chosen to satisfy conditions at infinity
{\rm{(\ref{kony_IDE_cond_inf_pr})}} (as the normalizing
condition); for $\widetilde C_0$ defined in this way, the restriction
$0<\varphi(u, \widetilde C_0)<1$ is valid for any finite
$u\in{\textbf{R}}_+$, i.e., for $\varphi(u)=\varphi(u,
\widetilde C_0)$, inequalities
{\rm(\ref{kony_IDE_restr_pr})} are fulfilled tacitly.
\item[3.] If the inequality \, $m(a-\lambda) + c\ge 0$\,
is fulfilled then the solution $\varphi(u)$ is concave on
${\textbf{R}_+}$; in particular this is true when
\begin{equation}\label{kony_dohod_pr}
c - \lambda m >0.
\end{equation}
\item[4.] If the inequality \, $m(a-\lambda) + c < 0$\,
 is true then $\varphi(u)$ is convex on a certain interval $[0, \widehat
u]$ where $\widehat u$ is an inflection point, $\widehat u>0$.
\item[5.] For small $u$, due to Lemma~\ref{l:1} above, the solution $\varphi(u)$
is represented by asymptotic power series
{\rm{(\ref{kony_asymp00_pr})-(\ref{kony_1D_k_pr})}}
where $C_0=\widetilde C_0$, $0<\widetilde C_0<1$.
\item[6.]  For large  $u$, the asymptotic
representation
\begin{equation}\label{kony_asympinfty_pr}
\varphi(u)=1 - K u^{1-2a/b^2} [1+o(1)], \qquad u\to\infty,
\end{equation}
takes place with $K=\widetilde C_0 \widetilde K>0$ where in
general the value $\widetilde K>0$ (as well as the value
$\widetilde C_0$) cannot be determined using local analysis
methods.
\end{enumerate}
\end{theorem}
\subsection{The "Degenerate" Problems and Their Exact Solutions}
\label{subsec:1.3}
A particular case of IDE (\ref{kony_IDE_pr})
is considered "degenerate" when some of its parameters are equal
to zero.
\subsubsection{ The First "Degenerate"  Case: $ \bf{a=b=0}$, \,
$ {\bf {\lambda >0}}$, \, $ {\bf m>0}$, \, $ {\bf c>\lambda m > 0
}$}
 \label{subsubsec:1.3.1}
For this case,  the "degenerate" IDE problem
\begin{equation}\label{kony_IDE_1_deg1_pr}
c \varphi^{\prime}(u) - \lambda [\varphi(u) - (J_m\varphi)(u)] = 0,
\qquad u\in{\textbf{R}}_+,
 \end{equation}
\begin{equation}\label{kony_IDE_cond_0_inf_deg1_pr}
c\varphi^\prime(0) - \lambda \varphi(0)=0, \qquad
\lim_{u\to\infty}\varphi(u)=1,
\end{equation}
is equivalent to the ODE problem with one parameter:
\begin{equation}\label{kony_diffur_deg1_pr}
c \varphi^{\prime\prime}(u) +
\left(c/m - \lambda\right)\varphi^\prime(u)=0,   \qquad
u\in{\textbf{R}}_+,
\end{equation}
\begin{equation}\label{kony_cond_0_C0_d_deg1_pr}
\varphi(0) = C_0,  \qquad
\varphi^{\prime}(0)=\lambda C_0/c,\qquad \lim_{u\to\infty}\varphi(u)=1.
\end{equation}
Then we obtain $C_0 = \widetilde C_0= 1 -\lambda m/c$,\,
 $0<\widetilde C_0<1$, and
\begin{equation}\label{kony_Pexp_pr}
\varphi(u)= \varphi(u,\widetilde C_0) = 1-\frac{\lambda m}{c}
\exp{ \left( - \frac{c - \lambda m}{mc} u \right)}, \qquad
u\in {\textbf{R}}_+.
\end{equation}

If inequality (\ref{kony_dohod_pr}) is not valid, i.e.,
$c\le\lambda m$, then there is no solution to problem
(\ref{kony_IDE_1_deg1_pr}), (\ref{kony_IDE_cond_0_inf_deg1_pr})
[resp., to problem (\ref{kony_diffur_deg1_pr}), (\ref{kony_cond_0_C0_d_deg1_pr})].

In what follows, function (\ref{kony_Pexp_pr}) is well known
in classical C.-L. risk theory  and has an independent
meaning (see further Sect.~3.1).
\subsubsection{The Second "Degenerate"  Case: ${\bf b=0}$, \,
${\bf a>0}$, \, ${\bf c\ge 0}$, \,  $\bf{ \lambda>0}$, \ ${ \bf
m>0}$}
\label{subsubsec:1.3.2}
For $c>0$,  the \,"degenerate"\, IDE problem
\begin{equation}\label{kony_IDE_1_deg2_pr_c}
\begin{array}{c}
(a u + c)\varphi^{\prime}(u) - \lambda[\varphi(u) -
(J_m\varphi)(u)] = 0, \qquad u\in{\textbf{R}}_+,
\\
\\
c\varphi^\prime(0) - \lambda\varphi(0)=0, \qquad
\lim_{u\to\infty}\varphi(u)=1,
\end{array}
\end{equation}
is equivalent to the parametrized ODE problem:
\begin{equation}\label{kony_diffur_deg2_pr_c}
\begin{array}{c}
(a u+c) \varphi^{\prime\prime}(u) + \left (a - \lambda + c/m +
au/m\right) \varphi^\prime(u)=0, \qquad u\in{\textbf{R}}_+,
\\
\\
\varphi(0) = C_0, \qquad \varphi^{\prime}(0) = \lambda C_0/c,
\qquad \lim_{u\to\infty}\varphi(u)=1.
\end{array}
\end{equation}
This implies $C_0 = \widetilde C_0=
(a/\lambda)(c/a)^{\lambda/a}\left [(a/\lambda)(c/a)^{\lambda/a}
+ I_c(0) \right]^{-1}$,\, $0<\widetilde C_0<1$,
\begin{equation}\label{kony_exp_deg2_pr_c}
\varphi(u)= \varphi(u,\widetilde C_0)= 1 - I_c(u)\left [I_c(0) +
(a/\lambda)(c/a)^{\lambda/a}\right]^{-1},  \quad u\in
{\textbf{R}}_+,
\end{equation}
where, taking into account the notation $\Gamma(p, z)=
\int_z^\infty{x^{p-1}\exp{(-x)}dx}$, \, $p>0$,  for incomplete
gamma-function (see, e.g., \cite{kony_BatErd}), we have
\begin{equation}\label{kony_int_deg2_c}
\begin{array}{c}
I_c(u) = \int_u^\infty{\left(x+c/a\right)^{\lambda/a-1}
\exp{(-x/m)}dx}=
\\
\\
= m^{\lambda/a}\exp{\Big(c/(am)\Big)}
\Gamma\Big(\lambda/a, \, u/m + c/(am) \Big), \quad u\ge 0.
\end{array}
\end{equation}
In particular we obtain the asymptotic representation when $u\to\infty$:
\begin{equation}\label{kony_asymp_deg2}
\varphi(u)= 1 - m\left[(a/\lambda)(c/a)^{\lambda/a} +
I_c(0)\right]^{-1} u^{\lambda/a-1}\exp{(-u/m)}[1+o(1)].
\end{equation}

For $c=0$, the solution to the IDE problem on
${\textbf{R}}_+$,
\begin{equation}\label{kony_IDE_1_deg2_pr_0}
u \varphi^{\prime}(u) - (\lambda/a) [\varphi(u) - (J_m\varphi)(u)] = 0,
\quad   \lim_{u\to +0}\varphi(u)= 0, \quad
\lim_{u\to\infty}\varphi(u)=1,
\end{equation}
can be found as a solution to the equivalent ODE problem:
\begin{equation}\label{kony_diffur_deg2_pr_0}
\begin{array}{c}
u^2\varphi^{\prime\prime}(u) + \left (1 - \lambda/a +
u/m\right) u \varphi^\prime(u)=0, \qquad u\in{\bf{R}}_+,
\\
\\
\lim_{u\to +0}\varphi(u)= \lim_{u\to +0}[u\varphi^\prime (u)]=0,
\qquad \lim_{u\to\infty}\varphi(u)=1.
\end{array}
\end{equation}
This implies  the same formulas
(\ref{kony_exp_deg2_pr_c})-(\ref{kony_asymp_deg2}) with $c=0$
where $\Gamma(p) = \Gamma(p,0)$ is the usual Euler
gamma-function. In particular, using the formula
$$\varphi^\prime(u) = [ m^{\lambda/a} \Gamma(\lambda/a)] ^{-1}
u^{\lambda/a-1} \exp{(-u/m)}, \quad u\ge 0,$$
 we obtain here: if
$a<\lambda$ then $\varphi^\prime(0)=0$; if $a=\lambda$ then
$\varphi^\prime(0)=1/m$ and $\varphi(u) = 1 - \exp{(-u/m)}$; if
$a>\lambda$ then the function $\varphi^\prime(u)$ is unbounded
as $u\to +0$ but integrable on ${\bf{R}}_+$.

This "degenerate" case has an independent meaning in risk
theory (see further Sect.~3.2).
\section{ Origin of the Problem: the Cram\'er-Lundberg Dynamic Insurance
Models}
\label{sec:2}
\subsection{ The Classical C.-L. Insurance Model}
\label{subsec:2.1}
 Consider the classical risk process:\,
$R_t=u+ct-\sum\limits_{k=1}^{N_t}Z_k, \, \, t\ge 0$. Here $R_t$
is the surplus of an insurance company at time $t$, $u$ is the
initial surplus, $c$ is the premium rate; $\{N_t\}$ is a Poisson
process with parameter $\lambda$ defining, for each $t$, the
number of claims applied on the interval $(0,t]$; $Z_1,
Z_2,\ldots $ is the series of independent identically
distributed random values with some distribution $F(z)$
($F(0)=0$, ${\bf{E}}Z_1=m<\infty$), describing the sequence of
claims; these random values are also assumed to be independent
of the process $\{N_t\}$. For this model, the positiveness
condition for the net expected income ("safety loading") has the
form (\ref{kony_dohod_pr}).

Denote by $\tau =\inf\{t:R_t < 0\}$ the time of ruin, then
$\bf{P}(\tau <\infty)$ is the probability of ruin at the
infinite time interval.

{\bf{A classical result in the C.-L. risk theory}}
\cite{kony_Grand}: under condition (\ref{kony_dohod_pr}) and
assuming existence of a constant $R_L>0$ ("the Lundberg
coefficient") such that equality $\int_0^\infty
[1-F(x)]\exp{(R_Lx)}dx=c/\lambda >0$ holds, the probability of
ruin $\xi(u)$ as a function of the initial surplus admits the
estimate \, $\xi(u)={\bf{P}}(\tau <\infty)\leq \exp{(-R_Lu)}$,
\, $u\ge 0$. Moreover, if the claims are exponentially
distributed,
\begin{equation}\label{kony_1F(x)}
F(x)=1-\exp{(-x/m)},\qquad m>0, \qquad x\ge 0,
\end{equation}
then $R_L=(c - \lambda m)/(mc)>0$, and the  survival probability
$\varphi(u)=1-\xi(u)$ is given by the exact formula
(\ref{kony_Pexp_pr}), i.e., coincides with the exact solution of
the first "degenerate" problem to which  input singular problem
(\ref{kony_IDE_pr})-(\ref{kony_IDE_cond_inf_pr}) reduces
formally as $a=b=0$ (see Sect.~\ref{subsubsec:1.3.1}).

For $c$ as a bifurcation parameter, the value $c=\lambda m$ is critical:
if $c\le\lambda m$ then
$\varphi(u)\equiv 0$,\, $u\in {\bf{R}}_+$.
\subsection{The C.-L. Insurance Model with
Investment into Risky Assets}
\label{subsec:2.2}
 Now consider the case where the surplus is invested continuously into shares
with price dynamics described by geometric Brownian motion
model:
\begin{equation}\label{kony_stock}
dS_t=S_t(a dt+b dw_t), \qquad t\ge 0.
\end{equation}
Here $S_t$ is the share price at time $t$, $a$ is the expected
return on shares, $0<b$ is the volatility, $\{w_t\}$ is a
standard Wiener process.

Denoting by $X_t $ the company's surplus at time  $t$ we get
$X_t =\theta _t S_t$, where  $\theta _t$ is the amount of shares
in the portfolio. Then the surplus dynamics meets the relation
$dX_t =\theta _tdS_t+dR_t.$
Taking into account (\ref{kony_stock}), we obtain:
\begin{equation}\label{kony_proc1}
dX_t=a X_tdt + b X_tdw_t+dR_t, \qquad t\ge 0.
\end{equation}
In contrast with the classical model, condition
(\ref{kony_dohod_pr}) (the positiveness of \, "safety loading")
is not assumed here.

For the dynamical process (\ref{kony_proc1}), the  survival
probability $\varphi (u)$ satisfies on ${\bf{R}}_+$ the
following linear IDE (see, e.g., \cite{kony_BelKonKur1}, \cite{kony_FKP} and references therein):
\begin{equation}
\label{kony_phi}
\lambda \int_0^u \varphi(u-z)dF(z) - \lambda \varphi(u) +
(au +c) \varphi^\prime (u) + (b^2/2) u^2
\varphi^{\prime\prime}(u)=0.
\end{equation}
From (\ref{kony_phi}), assuming exponential distribution of
claims (\ref{kony_1F(x)}) we get the initial IDE
(\ref{kony_IDE_pr}) under study.

Assuming that there exists the solution $\varphi(u)$ of IDE (\ref{kony_IDE_pr})
representing the survival probability as a function of
initial surplus, the following statement (further called FKP-theorem) was obtained
in \cite{kony_FKP}.
\begin{theorem}
\label{t:2}
Suppose $b>0$ and the claims are distributed exponentially, i.e.,
{\rm (\ref{kony_1F(x)})} is valid. Then:
\begin{enumerate}
\item[1.]
If  inequality (\ref{kony_r_0_pr1}) of "robustness of shares" is
fulfilled, then the asymptotic representation
(\ref{kony_asympinfty_pr}) holds with a certain constant $K>0$.
\item[2.] If \, $2a/b^2<1$, then $\varphi(u)\equiv 0$, $u\in {\bf{R}}_+$.
\end{enumerate}
\end{theorem}
 \subsection{The C.-L. Model with Investment into a Risk-Free Asset}
\label{subsec:2.3} The model under study comprises a more
general case where only a constant part   $\alpha$  ($0<\alpha<
1$)  of the surplus is invested in shares (with the expected
return $\mu$ and volatility $\sigma$) whereas remaining part
$1-\alpha$ is invested into a risk free asset (bank deposit with
constant interest rate $r>0$): the case $0<\alpha < 1$ may be
reduced to the case $\alpha=1$ by a simple change of the
parameters (shares characteristics), namely $a=\alpha \mu
+(1-\alpha)r$, \,$b=\alpha \sigma$.

Moreover, when  the surplus is invested entirely into a risk
free asset (bank deposit with constant interest rate), we obtain
the second "degenerate"  problem (with or without premiums) to
which the input singular problem
(\ref{kony_IDE_pr})-(\ref{kony_IDE_cond_inf_pr}) reduces
formally as $b=0$. For $a>0$, $\lambda>0$, $m>0$, $c\ge 0$,
there exists the exact solution (\ref{kony_exp_deg2_pr_c}),
(\ref{kony_int_deg2_c})
 and  the asymptotic representation (\ref{kony_asymp_deg2})
is valid (for details, see Sect.~\ref{subsubsec:1.3.2}).

Thus when the surplus is entirely invested into a risk free
asset then the survival probability is not equal to zero, for
$u>0$, even if premiums (insurance payments) are absent ($c=0$)
and has a good asymptotic behavior as $u\to\infty$.

The formulas (\ref{kony_exp_deg2_pr_c})-(\ref{kony_asymp_deg2}) see also in \cite{Paulsen Gj}.

\section{ On the Approach to Main Problem and Proofs of Main Results}
\label{sec:3}
\subsection{The Singular Problem for IDE:
Uniqueness of the Solution and Its Monotonic Behavior}
\label{subsec:3.1}
 As shown in Sect.~\ref{sec:2}, we can
formulate the input singular IDE problem in the form
(\ref{kony_1intdiff_exp_pr}), (\ref{kony_TrebDerivs_pr}),
(\ref{kony_IDE_cond_inf_pr1}), where operator $J_m$ is defined
by (\ref{int_exp_F_pr}), \, $C_0$ is an unknown parameter whose
value must be found, and, for the solution to the problem
(\ref{kony_1intdiff_exp_pr}), (\ref{kony_TrebDerivs_pr}),
(\ref{kony_IDE_cond_inf_pr1}), the restrictions needed are
(\ref{kony_IDE_restr_pr1}).
\begin{lemma}
\label{l:2}
For IDE {\rm (\ref{kony_1intdiff_exp_pr})}, let the
values $a$, $b$, $c$, $\lambda$ and $m$ be fixed with $c>0$,
$\lambda>0$, $m>0$ whereas $a$ and $b$ are any
real numbers ($a, b \in{\bf{R}}$). Then the following assertions
are valid:
\begin{enumerate}
\item[1.] If there exists a solution
$\varphi_1(u)=\varphi_1(u,C_0)$ to problem {\rm
(\ref{kony_1intdiff_exp_pr}), (\ref{kony_TrebDerivs_pr}),
(\ref{kony_IDE_cond_inf_pr1})} with some $C_0>0$, then it is a
unique solution to this problem.
\item[2.] Such $\varphi_1(u)$ satisfies restrictions {\rm (\ref{kony_IDE_restr_pr1})},
$0<C_0<1$ and $\varphi_1^\prime (u)>0$ for any finite $u\in
{\bf{R}_+}$, i.e., $\varphi_1(u)$ is a monotone nondecreasing on
${\bf{R}_+}$ function.
\end{enumerate}
\end{lemma}

{\it{Proof.}}
\begin{enumerate}
\item[1.] Supposing the opposite, let $\varphi_2(u)$ be any other solution
to problem  (\ref{kony_1intdiff_exp_pr}),
(\ref{kony_TrebDerivs_pr}), (\ref{kony_IDE_cond_inf_pr1}), i.e.,
$\varphi_2(u)\not \equiv\varphi_1(u)$. Then two cases may occur:
the first one with
\, $\lim_{u\to+0}\varphi_2(u)=\lim_{u\to+0}\varphi_1(u)$,\,  and  the
second one with
\, $\lim_{u\to+0}\varphi_2(u)\not=\lim_{u\to+0}\varphi_1(u)$.

\vspace{3mm}
For the first case,  it follows that there exists a nontrivial
solution $\widetilde{\varphi}(u)$ of IDE
(\ref{kony_1intdiff_exp_pr}) satisfying conditions \,
$\lim_{u\to+0}{\widetilde\varphi(u)}=
\lim_{u\to\infty}{\widetilde\varphi(u)}=0$.\, Let
$0<\widetilde{u}$ be its maximum point: \,
$\widetilde{\varphi}(\widetilde{u})=\max_{u\in
[0,\infty)}\widetilde{\varphi}(u)>0$\, (if
$\widetilde{\varphi}(u)$ takes only non-positive values then we
consider the solution $-\widetilde{\varphi}(u)$ instead). Then
\, $\widetilde{\varphi}^\prime(\widetilde{u})=0$, \,
$\widetilde{\varphi}^{\prime\prime}(\widetilde{u})\leq 0$.\, But
from IDE (\ref{kony_1intdiff_exp_pr}) a contradiction follows:
\begin{eqnarray*}
(b^2/2) \widetilde{u}^2 \widetilde{\varphi}^{\prime\prime}(\widetilde{u})&=&
\lambda [\widetilde\varphi(\widetilde{u}) -
m^{-1} \int_0^{\widetilde{u}}
{\widetilde{\varphi}(s) \exp{\big(-(\widetilde{u}-s)/m\big)} ds}] \\
&&
\ge\lambda \widetilde{\varphi}(\widetilde{u})\left [1 -
m^{-1}  \int_0^{\widetilde{u}}{\exp{\big(-(\widetilde{u}-s)/m\big)}
 ds}\right]
 \end{eqnarray*}
 \begin{equation}
  = \lambda\widetilde{\varphi}(\widetilde{u})\exp{(-\widetilde{u}/m)}>0.
 \end{equation}

For the second case, there exists a linear combination of
solutions \, $\widehat{\varphi}(u)=c_1\varphi_1(u) +
c_2\varphi_2(u)$\,  such that $\widehat{\varphi}(u)\not \equiv
1$ and satisfies conditions \, $\lim_{u\to+0}
{\widehat\varphi(u)} = \lim_{u\to\infty} {\widehat\varphi(u)} =
1$. \, If there exists a value $\widehat{u}>0$ with
$\widehat\varphi(\widehat u)>1$, then the first case argument is
valid. Otherwise, the inequality $\widehat\varphi(u)\le 1$ \,
$\forall u\in {\bf{R}_+}$ contradicts to \,
$\lim_{u\to+0}\widehat\varphi^\prime(u)=\lambda/c>0$\,  which
follows from (\ref{kony_TrebDerivs_pr}).
\item[2.] The other assertions are proved analogously.
\end{enumerate}
\subsection{SCPs for Accompanying Linear ODEs}
\label{sec:3.2}
\subsubsection{Reduction of the Second Order IDE to a Third-Order ODE}
\label{subsec:3.2.1}
The known possibility of reducing the
second order IDE (\ref{kony_1intdiff_exp_pr}) to a third order
ODE is important for further exposition. First, we note that
$$(J_m\varphi)^\prime (u)= \frac{1}{m}\left(\exp{(-u/m)}\int_0^u
\varphi(x)\exp{(x/m)}dx\right)^\prime $$
\begin{equation}\label{kony_rav}
=  [\varphi(u) - (J_m\varphi)(u)]/m.
\end{equation}
Then differentiating IDE (\ref{kony_1intdiff_exp_pr}) in view of
(\ref{kony_rav}) gives a linear third order IDE
\begin{equation}\label{kony_IDE_3}
\begin{array}{c}
(b^2/2) u^2 \varphi^{\prime\prime\prime}(u)+
[(b^2 + a) u+c] \varphi^{\prime\prime}(u) +
(a -\lambda)\varphi^\prime(u)+ \\
\\ +(\lambda/m)[\varphi(u) - (J_m\varphi)(u)]=0, \quad u\in {\bf{R}}_+,
\end{array}
\end{equation}
which also implies the limit condition
\begin{equation}\label{kony_diffur_cond1}
\lim_{u\to +0} [c\varphi^{\prime\prime}(u) + (a -\lambda) \varphi^\prime
(u) + (\lambda /m)\varphi(u)]= 0.
\end{equation}
Together with the input limit condition (\ref{kony_IDE_cond_0_1_pr})
it implies  the limit equality
\begin{equation}\label{kony_IDE_3_cond_0}
\lim_{u\to +0}[c \varphi^{\prime\prime}(u)+(a-\lambda
+c/m) \varphi^\prime (u)]=0.
\end{equation}
In order to remove the integral term, we add IDE
(\ref{kony_IDE_3}) and initial IDE (\ref{kony_1intdiff_exp_pr})
multiplied by $1/m$ and get the linear third order ODE
(\ref{kony_diffur_pr}). Then the same limit condition
(\ref{kony_IDE_3_cond_0}) must be fulfilled to provide a
degeneration of this ODE as $u\to +0$.

Suppose $\psi(u)=\varphi^\prime (u)$ and rewrite ODE
(\ref{kony_diffur_pr}) in more canonical forms for ODEs with
pole-type singularities at zero and infinity (for classification
of isolated singularities of linear ODE systems and general
theory of ODEs of this class, see, e.g., the monographs
\cite{kony_cod}, \cite{kony_Fedoryuk} and \cite{kony_Wasov}
complementing each other). Now, for $\psi(u)$, we have to study
the following singular ODEs: for small $u$, we need to consider
the equation
\begin{equation}\label{kony_diffur_psi_0}
\begin{array}{c}
(b^2/2) u^3 \psi^{\prime\prime}(u) + \left [c +
(b^2 + a) u +  b^2 u^2/(2m)\right]u \psi^{\prime}(u) + \\
\\ + \left [(a - \lambda + c/m) u + a u^2/m \right]\psi(u)=0, \qquad u>0,
\end{array}
\end{equation}
and for large $u$, we shall consider the same equation in the
form
\begin{equation}\label{kony_diffur_psi_inf}
\begin{array}{c}
(b^2/2) \psi^{\prime\prime}(u) + \left [c/u^2 +
(b^2 + a)/u +  b^2/(2m)\right]\psi^{\prime}(u) + \\
\\ + \left [(a - \lambda + c/m)/u^2 + (a/m)/u \right]\psi(u)=0,
\qquad u>0.
\end{array}
\end{equation}
We see that both ODE (\ref{kony_diffur_psi_0}) and equivalent
ODE (\ref{kony_diffur_psi_inf}) have irregular (strong)
singularities of rank $1$ as $u\to +0$ and as $u\to \infty$.
\subsubsection{ Singularity at Zero: Replacement of the SIP for IDE
by an Equivalent SCP for ODE}
\label{subsec:3.2.2}
\paragraph{\it Proof of Lemma~\ref{l:1}}
First, we must show that the previous transformations permit us
to replace the input SIP (\ref{kony_1intdiff_exp_pr}),
(\ref{kony_TrebDerivs_pr}) for an IDE by the  SCP
(\ref{kony_diffur_pr}), (\ref{kony_cond_0_C0_d_pr}) for an ODE.

In the straight direction (from the IDE SIP  to the ODE SCP),
the statement is evident. Now let
$\widetilde\varphi(u)=\widetilde\varphi(u,C_0)$ be a solution of
ODE SCP (\ref{kony_diffur_pr}), (\ref{kony_cond_0_C0_d_pr}). We
have to prove that $\widetilde\varphi(u)$ satisfies IDE
(\ref{kony_1intdiff_exp_pr}).

Denote the left part of IDE (\ref{kony_1intdiff_exp_pr}) with the function
$\widetilde\varphi(u)$ by $g(u)$. We have to prove that $g(u)\equiv 0$. Indeed,
the way ODE (\ref{kony_diffur_pr}) was derived means that $g(u)$ meets the first-order ODE
$$g^\prime (u) + g(u)/m=0, \qquad 0\le u <\infty,$$
with the general solution of the form $g(u)= \widetilde C
\exp(-u/m)$ where $\widetilde C$ is an arbitrary constant. Since
$\widetilde\varphi(u,C_0)$ meets conditions
(\ref{kony_cond_0_C0_d_pr}), it follows from IDE
(\ref{kony_1intdiff_exp_pr}) that $g(0)=0$. This implies
$\widetilde C=0$, i.e., $g(u)\equiv 0$.

The other statements of Lemma~\ref{l:1} follow from the results
of \cite{kony_Kony1} (see \cite{kony_BelKonKur2} for details).
\subsubsection{ SCP at Infinity and Its Two-Parameter Family of Solutions}
\label{subsubsec:3.2.3}
 For $\psi(u)=\varphi^\prime(u)$, we have an SCP at infinity for the second order ODE
(\ref{kony_diffur_psi_inf}) with the conditions
\begin{equation}\label{kony_1cond_infty}\lim_{u\to\infty} \psi(u)=
\lim_{u\to\infty}\psi^\prime(u)=0.
\end{equation}
Using the known results for linear ODEs  with irregular
singularities, we obtain the following assertions (more complete
in comparison with FKP-theorem).
\begin{lemma}
\label{l:3} For ODE {\rm{(\ref{kony_diffur_psi_inf})}}, suppose
that $b\not= 0$, $a>0$, $m>0$ whereas  $\lambda$ and $c$ are
arbitrary real numbers ($\lambda, c \in {\bf{R}}$). Then:
\begin{enumerate}
\item[1.]
Any solution to ODE {\rm{(\ref{kony_diffur_psi_inf})}} satisfies
conditions {\rm{(\ref{kony_1cond_infty})}} so that SCP
{\rm{(\ref{kony_diffur_psi_inf})}},
{\rm{(\ref{kony_1cond_infty})}} at infinity has a two-parameter
family of solutions $\psi(u, d_1, d_2)$ where $d_1$ and $d_2$
are arbitrary constants.
\item[2.] For this family, the following representation holds:
\begin{equation}\label{kony_1psi_asimp}
\begin{array}{c}
\psi(u,d_1,d_2)= d_1  u^{-2a/b^2} [1+\chi_1(u)/u] +
\\
\\
+ d_2 u^{-2} \exp{(-u/m)} [1+\chi_2(u)/u)];
\end{array}
\end{equation}
here the functions $\chi_j(u)$ have finite limits as
$u\to\infty$ and, for large $u$, can be represented by
asymptotic series in inverse integer powers of $u$,
\begin{equation}\label{expansion_for_psi}
\chi_j(u)\sim\sum_{k=0}^\infty \chi_j^{(k)}/u^k, \qquad j=1,2,
\end{equation}
where the coefficients $\chi_j^{(k)}$ may be found by
substitution of {\rm(\ref{kony_1psi_asimp}),
(\ref{expansion_for_psi})} in ODE
{\rm(\ref{kony_diffur_psi_inf})} ($j=1,2$, \, $k\ge 0$).
\item[3.] All solutions of the family {\rm(\ref{kony_1psi_asimp})}
are integrable at infinity
iff inequality {\rm(\ref{kony_r_0_pr1})} is fulfilled.
\end{enumerate}
\end{lemma}
For a detailed proof of Lemma~\ref{l:3}, see
\cite{kony_BelKonKur2}.
\begin{corollary}
\label{c:1} Under the assumptions of Lemma~\ref{l:3}, all
solutions of ODE {\rm(\ref{kony_diffur_pr})} have finite limits
as $u\to\infty$ iff condition {\rm(\ref{kony_r_0_pr1})}  is
fulfilled.
\end{corollary}

Summarizing all results, we obtain the proof of
Theorem~\ref{t:1}.
\section{ The Accompanying Singular Problem
for Capital Stock Model (the Third "Degenerate" Case: $\bf
{c=0}$, $\bf{b\not=0}$, $\bf{a>0}$, $\bf{\lambda>0}$,
$\bf{m>0}$)} \label{sec:4} For this case, the input singular IDE
problem has the form:
 \begin{equation}\label{kony_IDE_c0}
(b^2/2) u^2 \varphi^{\prime\prime}(u) + au \varphi^{\prime}(u) -
\lambda [\varphi(u) - (J_m \varphi)(u)]=0, \quad
u\in\textbf{R}_+,
\end{equation}
\begin{equation}\label{kony_IDE_cond_0_c0}
\lim_{u\to+0}{\varphi(u)}=\lim_{u\to+0}
{[u\varphi^{\prime}(u)]}=0,
\end{equation}
\begin{equation}\label{kony_IDE_cond_inf_c0}
\lim_{u\to\infty}\varphi(u)=1, \qquad \lim_{u\to\infty}
\varphi^\prime(u)=0,
\end{equation}
and restrictions (\ref{kony_IDE_restr_pr}) are needed for the
solution.

The following lemma is analogous to Lemma~\ref{l:2} (with a
similar proof).
\begin{lemma}
\label{l:4}
For IDE {\rm (\ref{kony_IDE_c0})}, let the values
$a$, $b$, $\lambda$ and $m$ be fixed with $\lambda>0$,
$m>0$ whereas $a$ and $b$ are any real numbers ($ a,b
\in\bf{R}$). Then the following assertions are valid:
\begin{enumerate}
\item[1.]
If there exists a solution $\varphi_1(u)$ to the problem
{\rm(\ref{kony_IDE_c0})-(\ref{kony_IDE_cond_inf_c0})}, then it
is a unique solution to this problem.
\item[2.]
Such $\varphi_1(u)$ satisfies restrictions {\rm
(\ref{kony_IDE_restr_pr})}  and $\varphi_1^\prime (u)>0$ for any
finite $u>0$, i.e., $\varphi_1(u)$ is a monotone nondecreasing
on $\bf{R_+}$ function.
\end{enumerate}
\end{lemma}

Analogously to the previous approach, the singular IDE problem
(\ref{kony_IDE_c0})-(\ref{kony_IDE_cond_inf_c0}) is equivalent
to the following singular ODE problem:
\begin{equation}\label{kony_diffur_c0}
\begin{array}{c}
(b^2/2) u^3 \varphi^{\prime\prime\prime}(u) +
\left [b^2 + a +  b^2 u/(2m)\right] u^2\varphi^{\prime\prime}(u) + \\
\\ + \left (a - \lambda + a u/m \right)u \varphi^\prime(u)=0,
\quad 0<u<\infty,
\end{array}
\end{equation}
\begin{equation}\label{kony_cond_0_c0}
\lim_{u\to+0}{\varphi(u)} = \lim_{u\to+0}{[u
\varphi^{\prime}(u)]}=\lim_{u\to+0}{[u^2
\varphi^{\prime\prime}(u)]}=0,
\end{equation}
\begin{equation}\label{kony_cond_inf_d_c0}
\lim_{u\to\infty}\varphi(u)=1, \qquad \lim_{u\to\infty}
\varphi^\prime(u)=\lim_{u\to\infty} \varphi^{\prime\prime}(u)=0.
\end{equation}

First, consider SCP at regular (weak) singular point $u=0$,
i.e., SCP (\ref{kony_diffur_c0}), (\ref{kony_cond_0_c0})
introducing notation:
\begin{equation}\label{kony_mu1_c0}
\mu_1 = 1/2 - a/b^2 + \sqrt{(1/2 - a/b^2)^2 + 2\lambda/b^2},
\end{equation}
\begin{equation}\label{kony_eta_d1_d2_c0}
d_1=\mu_1 + a/b^2, \qquad d_2=\mu_1 + 2a/b^2 - 1.
\end{equation}
The following lemma is analogous to Lemma~\ref{l:1}.
\begin{lemma}
\label{l:4}
For IDE {\rm(\ref{kony_IDE_c0})}, let the values
$a$, $b$, $\lambda$, $m$ be fixed with $b\not =0$, $\lambda
>0$, $m>0$, $a\in \bf{R}$. Then:
\begin{enumerate}
\item[1.]
The IDE SIP {\rm(\ref{kony_IDE_c0}), (\ref{kony_IDE_cond_0_c0})}
is equivalent to the ODE SCP {\rm(\ref{kony_diffur_c0}),
(\ref{kony_cond_0_c0})}.
\item[2.]
There exists a one-parameter family of solutions $\varphi(u,
P_1)$ to the ODE SCP {\rm{(\ref{kony_diffur_c0}),
(\ref{kony_cond_0_c0})}} (therefore also to the equivalent IDE
SIP {\rm(\ref{kony_IDE_c0})}, {\rm(\ref{kony_IDE_cond_0_c0})})
and the following representation holds:
\begin{equation}\label{kony_eta_repr_c0}
\varphi(u,P_1) = P_1  \int_0^u {s^{\mu_1-1} \eta(s) ds};
\end{equation}
here $P_1$ is a parameter, \,$0<\mu_1$ is defined by
{\rm(\ref{kony_mu1_c0})},  and $\eta(u)$ is a solution to SCP
\begin{equation}\label{kony_eta_eq_c0}
u^2 \eta^{\prime\prime}(u) + (2d_1 + u/m) u\eta^{\prime}(u)
+ (d_2 u/m) \eta(u)=0, \quad u>0,
\end{equation}
\begin{equation}\label{kony_eta_cond_0_c0}
\lim_{u\to+0}{\eta(u)} = 1, \qquad \lim_{u\to+0}{[u
\eta^{\prime}(u)]}=0,
\end{equation}
where $d_1$ and $d_2$ are defined by
{\rm(\ref{kony_eta_d1_d2_c0})}; there exists a unique solution
$\eta(u)$ to the SCP {\rm{(\ref{kony_eta_eq_c0}),
(\ref{kony_eta_cond_0_c0})}} and it is a holomorphic function at
the point $u=0$,
\begin{equation}\label{kony_eta_ser_c0}
\eta(u)= 1 + \sum_{k=1}^{\infty}P_{k+1}u^k, \quad |u|\le u_0,
\quad u_0>0,
\end{equation}
where the coefficients $P_{k+1}$ may be found by formal
substitution of series {\rm(\ref{kony_eta_ser_c0})} into ODE
{\rm(\ref{kony_eta_eq_c0})}, namely, from the recurrence
relations:
\begin{equation}\label{kony_1D_2_c0}
P_2=-d_2/(2md_1),
\end{equation}
\begin{equation}\label{kony_1D_k_c0}
P_{k+1} = - P_k (k - 1 + d_2)/[mk(k - 1 + 2 d_1)], \quad k=2, 3,
\ldots;
\end{equation}
moreover, if $D_1=\lim_{u\to+0}\varphi^\prime(u, P_1)$, then
$D_1=0$ when $a<\lambda$;  $D_1=P_1$ when $a=\lambda$; and at last
 $|D_1|=\infty$ when $a>\lambda$ (but $\varphi^\prime(u, P_1)$
is integrable as $u\to +0$).
\end{enumerate}
\end{lemma}

Summarizing the results and taking into account that
Lemma~\ref{l:3} and Corollary~\ref{c:1} are valid for any
$c\in\bf{R}$, we obtain
\begin{theorem}
\label{t:3}
 For IDE {\rm{(\ref{kony_IDE_c0})}}, let all the
parameters $a$, $b$, $\lambda$, $m$ be fixed positive numbers
and let inequality {\rm(\ref{kony_r_0_pr1})} of "robustness
of shares" be fulfilled. Then the following assertions are
valid:
\begin{enumerate}
\item[1.]
There exists a unique solution $\varphi(u)$ of singular linear
IDE problem
{\rm{(\ref{kony_IDE_c0})-(\ref{kony_IDE_cond_inf_c0})}}, it
satisfies restrictions {\rm (\ref{kony_IDE_restr_pr})} and, for
$u>0$, is a smooth monotone nondecreasing function.
\item[2.]
Such $\varphi(u)$ can be obtained by the formula
\begin{equation}\label{kony_eta_repr_fin_c0}
\varphi(u) = \int_0^u {s^{\mu_1-1} \eta(s)
ds}\Big/\int_0^\infty {s^{\mu_1-1} \eta(s) ds}, \qquad u\ge
0,
\end{equation}
where $\eta(u)$ is defined in Lemma~\ref{l:4}.
\item[3.]
For finite $u>0$, the solution $\varphi(u)$ is represented by
a convergent series which can be obtained using formulas
{\rm{(\ref{kony_eta_repr_fin_c0}),
(\ref{kony_eta_ser_c0})}-(\ref{kony_1D_k_c0})}.
\item[4.]
If  $a\ge\lambda$ then the solution $\varphi(u)$ is concave
on $\bf{R_+}$; moreover, if   $a=\lambda$  then
$\lim_{u\to+0}\varphi^\prime(u)=1/\int_0^\infty{\eta(s)ds}>0$, and if
$a>\lambda$ then $\lim_{u\to+0}\varphi^\prime(u)=\infty$ but
$\varphi^\prime(u)$ is an integrable on $\bf{R}_+$ function.
\item[5.]
If  $a<\lambda$ then $\lim_{u\to+0}\varphi^\prime(u)=0$, and $\varphi(u)$ is convex on a
certain interval $[0, \widehat u]$ where $\widehat u$ is an
inflection point, $\widehat u>0$.
\item[6.]
For large  $u$, the asymptotic representation
{\rm(\ref{kony_asympinfty_pr})} holds with $K>0$ where in
general the value $K>0$ cannot be determined using local
analysis methods.
\end{enumerate}
\end{theorem}
\section{Numerical Examples and Their Interpretation}
\label{sec:5}
For the main case $c>0$, our study shows that
the input singular IDE problem
(\ref{kony_IDE_pr})-(\ref{kony_IDE_cond_inf_pr}) may be reduced
to the auxiliary ODE SCP (\ref{kony_diffur_pr}),
(\ref{kony_cond_0_C0_d_pr}) with the parameter $C_0$ to be
defined, $0<C_0<1$. The asymptotic  expansion of the solutions
at zero (\ref{kony_asymp00_pr}) is used to transfer the limit
initial conditions  (\ref{kony_cond_0_C0_d_pr}) from the
singular point $u=0$ to a nearby regular point $u_0>0$; the
derivatives of the solution may be evaluated by formal
differentiation of the representation (\ref{kony_asymp00_pr}).
Consequently, a regular Cauchy problem  is to be solved starting
from the point $u_0>0$. The parameter $C_0$ in
(\ref{kony_asymp00_pr}) is evaluated numerically to satisfy
 the condition $\lim_{u\to\infty}\varphi(u)=1$.

For the additional case $c=0$, the singular IDE problem
(\ref{kony_IDE_c0})-(\ref{kony_IDE_cond_inf_c0}) is equivalent
to the singular ODE problem
(\ref{kony_diffur_c0})-(\ref{kony_cond_inf_d_c0}). To solve this
problem we use formula (\ref{kony_eta_repr_fin_c0}) and the
auxiliary SCP (\ref{kony_eta_eq_c0}),
(\ref{kony_eta_cond_0_c0}). The convergent power series
(\ref{kony_eta_ser_c0})-(\ref{kony_1D_k_c0}) is used to transfer
limit initial conditions (\ref{kony_eta_cond_0_c0}) from the
singular point $u=0$ to a regular point $u_0>0$, and then a
regular Cauchy problem is to be solved starting from this point.

Maple programming package was used as a numerical tool.

For all examples, we put $m=1$,  $\lambda=0.09$, and,
for $a>0$, $b\not= 0$, the shares are "robust": $2a/b^2>1$ (Figs.1--5).
\begin{figure}[t]
\centering
\includegraphics[scale=.35]{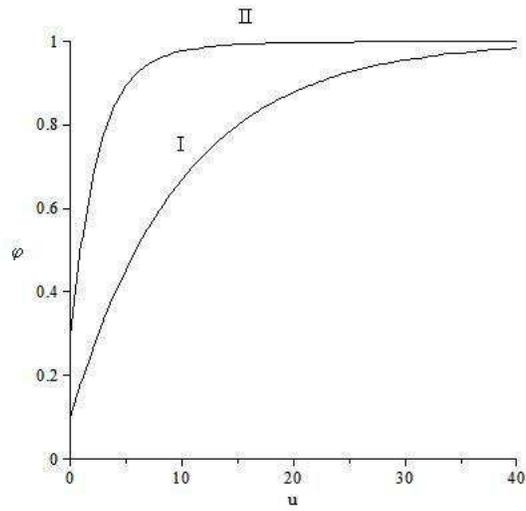}
\caption{The case ${\bf {c > \lambda m}:}$ $c=0.1$;  {\bf I:}\,
${\bf a=b=0}$  (the first "degenerate" case with the exact
solution); $C_0 =\varphi(0)=0.1$, $D_1=\varphi^\prime(+0)=0.09$;
{\bf II:}\, ${\bf a=0.02}$, ${\bf b=0.1}$; $C_0=0.295$,
$D_1=0.265$} \label{fig:1}
\end{figure}
\begin{figure}[t]
\centering
\includegraphics[scale=.35]{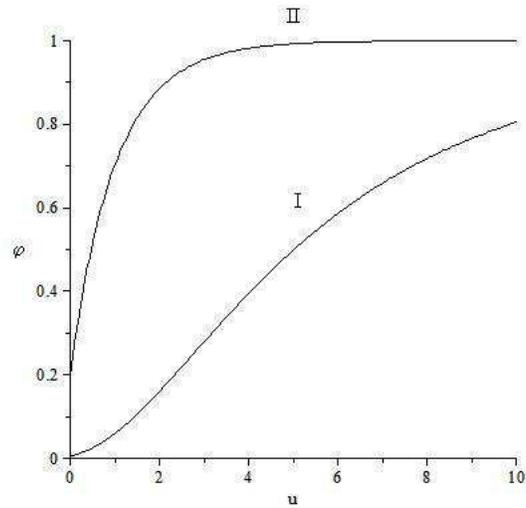}
\caption{The case ${\bf {c < \lambda\,m}:}$ $c=0.02$, ${\bf
b=0.1}$; {\bf I:}\, ${\bf a=0.02}$ (${\bf {m (\lambda - a)>
c}}$: $\varphi(u)$ has an inflection); $C_0 = 0.00527$, $D_1 =
0.0237$; {\bf II:} \, ${\bf a=0.1}$ (${\bf{m (\lambda - a)<
c}}$: $\varphi(u)$ is  concave); $C_0=0.194$, $D_1=0.872$}
\label{fig:2}
\end{figure}

\begin{figure}[t]
\centering
\includegraphics[scale=.35]{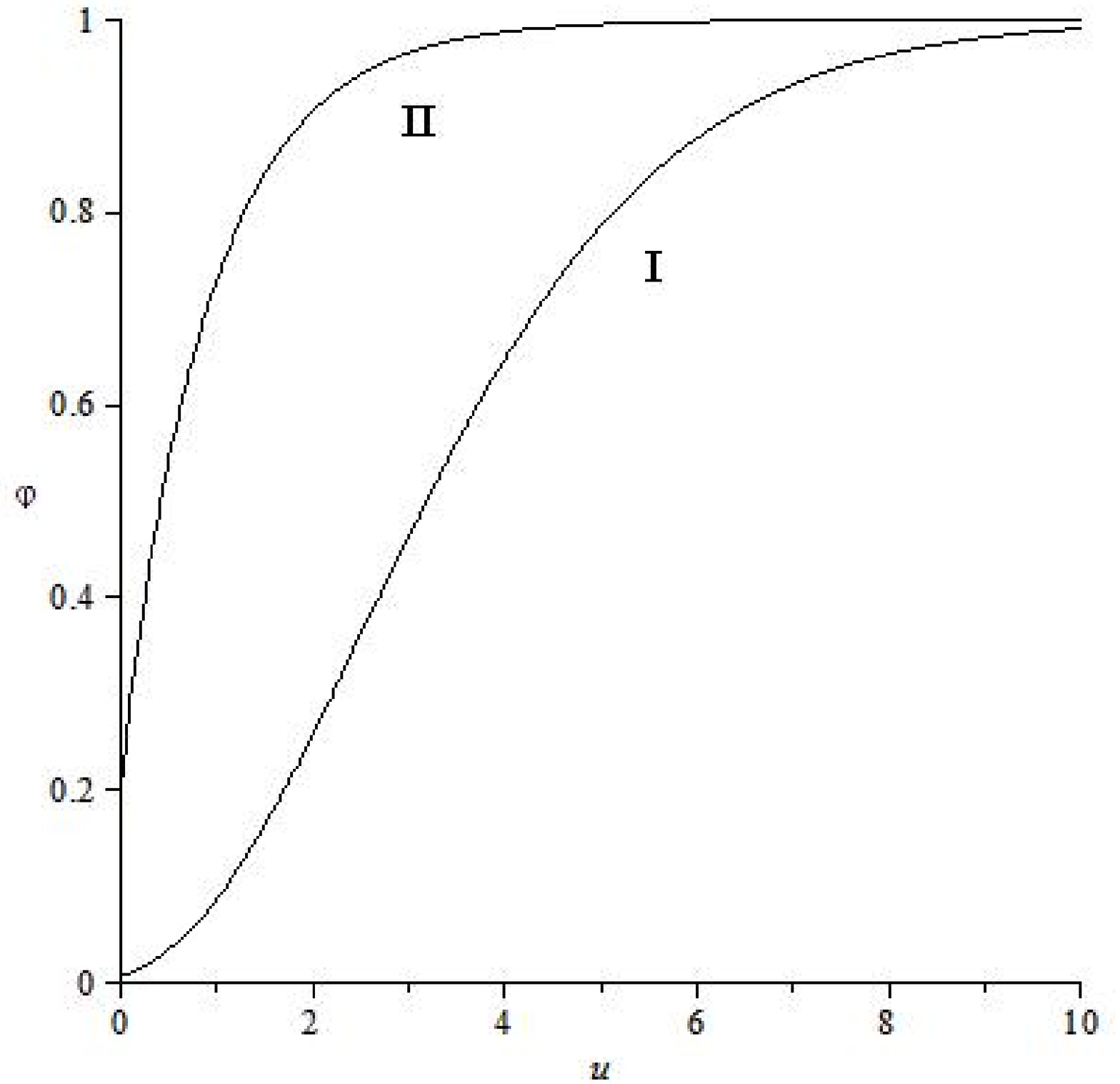}
\caption{The second "degenerate" case with premiums: ${\bf
b=0}$, $c=0.02$ (${\bf {c < \lambda\,m}})$; {\bf I:}\, ${\bf
a=0.02}$ (${\bf {m (\lambda - a)> c}}$); $C_0 = 0.00704$, $D_1 =
0.0317$; {\bf II:}\,  ${\bf a=0.1}$ (${\bf{m (\lambda - a)<
c}}$); $C_0=0.2046$, $D_1=0.9207$}
\label{fig:3}
\end{figure}
\begin{figure}[t]
\centering
\includegraphics[scale=.35]{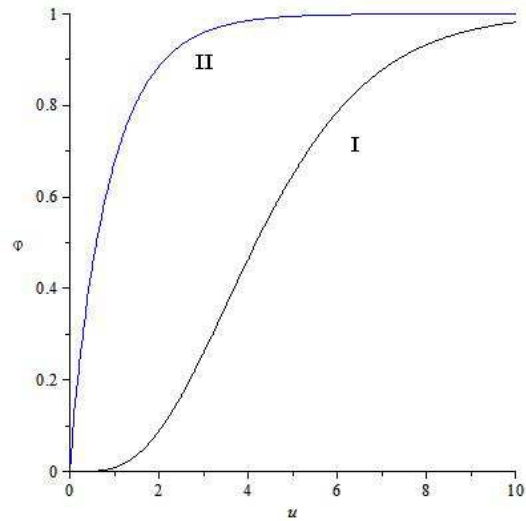}
\caption{The second "degenerate" case without premiums: ${\bf
b=0,\, c=0}$;   {\bf I:}\, ${\bf a=0.02}$ (${\bf {\lambda >a}}$);
$\varphi (0) = \varphi^\prime(0) = 0$; {\bf II:}
\,  ${\bf a=0.1}$ (${\bf{\lambda< a}}$); $\varphi (0) =
0$, $\varphi^\prime(+0) =\infty$}
\label{fig:4}
\end{figure}
\begin{figure}[t]
\centering
\includegraphics[scale=.35]{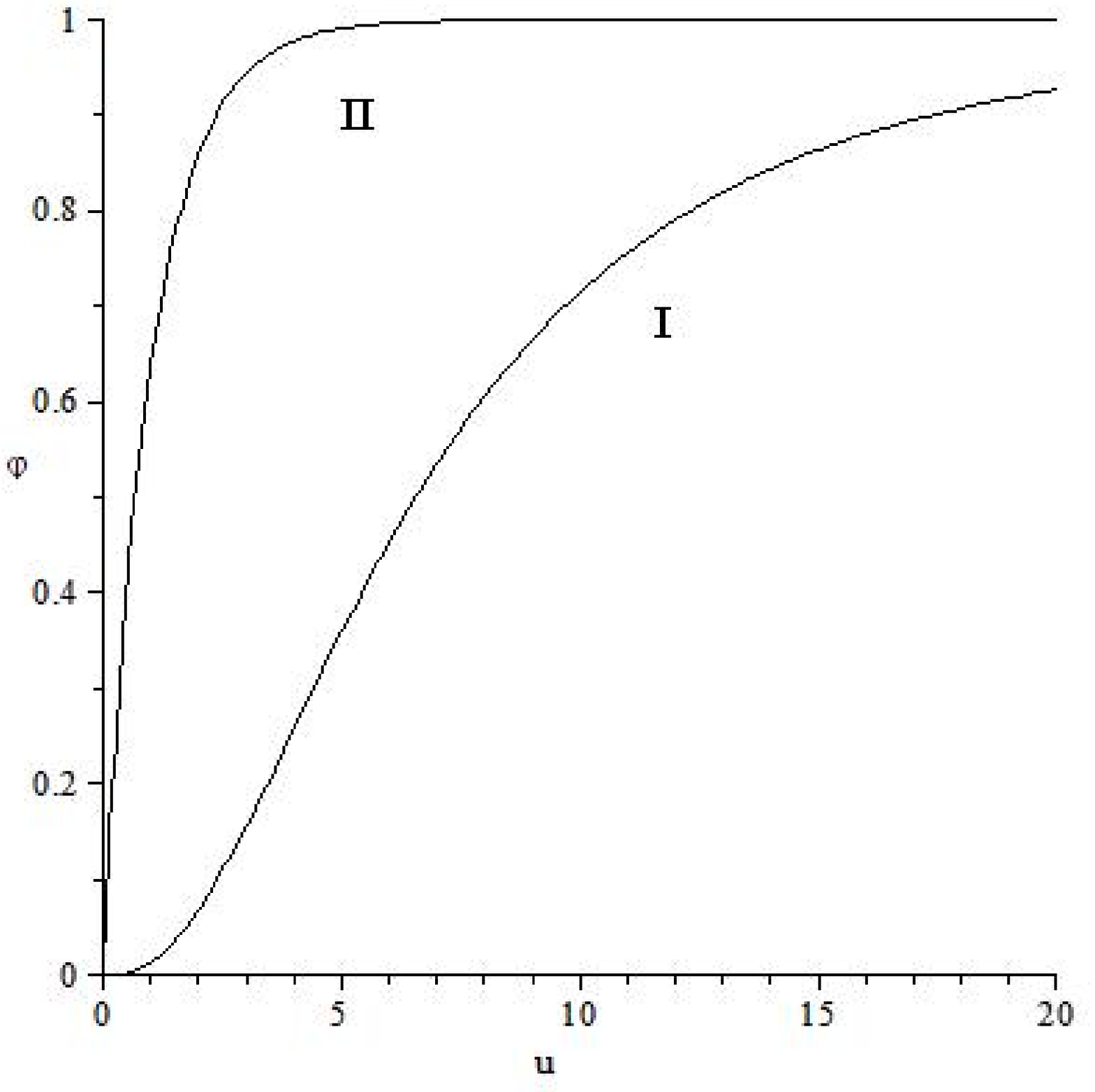}
\caption{The third "degenerate" case (the capital stock model):
${\bf c=0}$, ${\bf b=0.1}$; {\bf I:}\, ${\bf a=0.02}$ (${\bf
{\lambda>a}}$); $\varphi (0) = \varphi^\prime(0) = 0$,
$P_1=0.059587$; {\bf II:} \, ${\bf a=0.1}$ (${\bf{\lambda<a}}$);
$\varphi (0) = 0$, $\varphi^\prime(+0) =\infty$, $P_1=0.861816$
}
\label{fig:5}
\end{figure}
 \section{Conclusions}
The study shows that use of risky assets is not favorable for
non-ruin with large initial surplus values and constant structure of the
portfolio. However, the study of the cases when positiveness of the
safety loading does not hold shows risky assets to be effective
for small initial surplus values: while ruin is inevitable in the case without
investing, the survival probability grows considerably as $u$ grows
in presence of investing even if the premiums are absent
(moreover, the second derivative of the solution
for small $u$ is positive!). The study in \cite{kony_BelKonKur1}, \cite{kony_BelKonKur2} of
the optimal strategy for exponential distribution of claims
shows that the part of risky investments should be  $O(1/x)$ as
present surplus $x$ tends to infinity.
\begin{acknowledgement}
This work was supported by the  Russian Fund for Basic Research:
Grants RFBR 10-01-00767 and RFBR 11-01-00219.
\end{acknowledgement}


\begin{thebibliography}{99}

\bibitem{kony_AzMul}
Azcue, P., Muler, N.: Optimal investment strategy to minimize
the ruin probability of an insurance company under borrowng
constraints. Insurance Math. Econom. \textbf{44}(1) 26--34
(2009)

\bibitem{kony_BatErd} Bateman, H., Erd\'elyi, A.: Higher
Transcendental Functions. McGraw-Hill, New York (1953)

\bibitem{kony_BelKonKur1}
Belkina, T.A., Konyukhova, N.B.,  Kurkina, A.O.:  Optimal
investment problem in the dynamic insurance models: I.
Investment strategies and the ruin probability. Survey on
Applied and Industrial Mathematics. \textbf{16}(6), 961--981
(2009) [in Russian]

\bibitem{kony_BelKonKur2}
Belkina, T.A., Konyukhova, N.B.,  Kurkina, A.O.:  Optimal
investment problem in the dynamic insurance models: II.
Cram\'er-Lundberg model with the exponential claims. Survey on
Applied and Industrial Mathematics. \textbf{17}(1),  3--24
(2010) [in Russian]

\bibitem{kony_cod}
Coddington,  E.A., Levinson, N.: Theory of Ordinary
Differential Equations. McGraw-Hill, New York (1955)

\bibitem{kony_Fedoryuk}
Fedoryuk, M.V.: Asymptotic Analysis: Linear
Ordinary Differential Equations. Springer, Berlin (1993)

\bibitem{kony_FKP}
Frolova, A., Kabanov, Yu., Pergamenshchikov, S.: In the
insurance business risky investments are dangerous. Finance
Stochast. \textbf{6}(2), 227--235 (2002)

\bibitem{kony_Grand}
Grandell, J.: Aspects of Risk Theory.  Springer, Berlin (1991)

\bibitem{kony_Kony1}
Konyukhova, N.B.: Singular Cauchy problems
for systems of ordinary differential equations. U.S.S.R. Comput.
Maths. Math. Phys.    \textbf{23}(3), 72--82 (1983)

\bibitem{Paulsen Gj}
Paulsen, J. and  Gjessing, H.K.: Ruin theory with stochastic return on investments.
Adv.  Appl. Probab.  \textbf{29}(4), 965--985 (1997)

\bibitem{kony_Wasov}
Wasov, W.: Asymptotic Expansions for Ordinary Differential Equations.
Dover, New York (1987)
\end{thebibliography}
\end{document}